\newif\ifcds\cdstrue                       
\newif\ifplb\plbfalse                      
\newif\ifjhep\jhepfalse                    
\ifcsname useelstitle\endcsname\plbtrue\fi 
\ifcsname jname\endcsname\jheptrue\fi      
\ifplb\cdsfalse\fi                         
\ifjhep\cdsfalse\fi                        
\ifcds                                     
\documentclass[ALICE,manyauthors]{cernphprep}
\fi
\usepackage{units}
\usepackage[shortcuts]{extdash}
\usepackage{lineno} 
\usepackage{xcolor}
\usepackage[textsize=tiny]{todonotes}
\usepackage{hyperref}
\usepackage[T1]{fontenc}
\usepackage{orcidlink}

\definecolor{alicered}{HTML}{E30613}    
\definecolor{aliceyellow}{HTML}{EF7B0B} 
\definecolor{alicepurple}{HTML}{CC4645} 
\definecolor{alicegray}{HTML}{2F3C47}   
\definecolor{cernblue}{HTML}{0053A1}
\newcommand{\Npart}        {\ensuremath{N_\mathrm{part}}\xspace}

\newcommand{\nineH}        {$\sqrt{s}~=~0.9$~Te\kern-.1emV\xspace}
\newcommand{\seven}        {$\sqrt{s}~=~7$~Te\kern-.1emV\xspace}
\newcommand{\twoH}         {$\sqrt{s}~=~0.2$~Te\kern-.1emV\xspace}
\newcommand{\twosevensix}  {$\sqrt{s}~=~2.76$~Te\kern-.1emV\xspace}
\newcommand{\five}         {$\sqrt{s}~=~5.02$~Te\kern-.1emV\xspace}
\newcommand{\twosevensixnn}{$\sqrt{s_{\mathrm{NN}}}~=~2.76$~Te\kern-.1emV\xspace}
\newcommand{\fivenn}       {$\sqrt{s_{\mathrm{NN}}}~=~5.02$~Te\kern-.1emV\xspace}

\newcommand{\GeVc}         {Ge\kern-.1emV/$c$\xspace}
\newcommand{\MeVc}         {Me\kern-.1emV/$c$\xspace}
\newcommand{\TeV}          {Te\kern-.1emV\xspace}
\newcommand{\GeV}          {Ge\kern-.1emV\xspace}
\newcommand{\MeV}          {Me\kern-.1emV\xspace}
\newcommand{\GeVmass}      {Ge\kern-.2emV/$c^2$\xspace}
\newcommand{\MeVmass}      {Me\kern-.2emV/$c^2$\xspace}

\newcommand{\VZERO}        {\rm{V0}\xspace}

\newcommand\TeVBF{\ensuremath\mathbf{Te\kern-.1emV\xspace}}
\newcommand\FiveTeV{\ensuremath%
  \sqrt{s_{\scriptscriptstyle\mathrm{NN}}}=\unit[5.02]{\TeV}}
\ifplb
\let\FiveTeVBF\FiveTeV
\else
\newcommand\FiveTeVBF{\mathbf{\ensuremath%
    \sqrt{\boldsymbol{s}_{\scriptscriptstyle\mathbf{NN}}}=\unit[5.02]{\TeVBF}}}
\fi
\newcommand\TwoTeV{\ensuremath%
  \sqrt{s_{\scriptscriptstyle\mathrm{NN}}}=\unit[2.76]{\TeV}}
\newcommand\textdndeta[1][\mathrm{ch}]{%
  \ensuremath\mathrm{d}N_{#1}/\mathrm{d}\eta}
\newcommand\etalab{\ensuremath\eta_{\mathrm{lab}}}
\newcommand\textdndetalab[1][\mathrm{ch}]{%
  \ensuremath\mathrm{d}N_{#1}/\mathrm{d}\etalab}
\newcommand\textdndy[1][\mathrm{ch}]{%
  \ensuremath\mathrm{d}N_{#1}/\mathrm{d}y}
\newcommand\textdEdy{%
  \ensuremath\mathrm{d}E_{\scriptscriptstyle\mathrm{T}}/\mathrm{d}y}
\newcommand\dispdndeta[1][\mathrm{ch}]{%
  \ensuremath\frac{\mathrm{d}N_{#1}}{\mathrm{d}\eta}}
\newcommand\dispdndy[1][\mathrm{ch}]{%
  \ensuremath\frac{\mathrm{d}N_{#1}}{\mathrm{d}y}}
\newcommand\dispdEdy{%
  \ensuremath\frac{\mathrm{d}E_{\scriptscriptstyle\mathrm{T}}}{\mathrm{d}y}}
\newcommand\INELGtZero{\mbox{INEL\textgreater0}}
\newcommand\ST{\ensuremath S_{\mathrm{T}}}
\newcommand\Ebj{\ensuremath \varepsilon_{\mathrm{Bj}}}
\newcommand\Elb{\ensuremath \varepsilon_{\mathrm{LB}}}
\newcommand\mT{\ensuremath m_{\scriptscriptstyle\mathrm{T}}}
\newcommand\pT{\ensuremath p_{\scriptscriptstyle\mathrm{T}}}

\newcommand\textpTm{\ensuremath \pT/m}

\newcommand\textavgpTm{\ensuremath \langle\pT\rangle/\langle m\rangle}
\newcommand\dispavgpTm{\ensuremath \frac{\langle\pT\rangle}{\langle
    m\rangle}}
\newcommand\fm{\ensuremath \mathrm{fm}}
\newcommand\GeVfmc{\ensuremath\GeV\!/(\fm^2\mathit{c})}
\newcommand\GeVfm{\ensuremath\GeV\!/\fm^3}
\renewcommand\GeVc{\ensuremath\GeV\!/\mathit{c}^2}
\renewcommand\MeVc{\ensuremath\MeV\!/\mathit{c}}
\newcommand\figref[1]{Fig.~\ref{#1}}
\newcommand\figureref[1]{\figurename~\ref{#1}}
\newcommand\EqRef[1]{Eq.~\eqref{#1}}
\newcommand\sQgp{sQGP}
\renewcommand\VZERO[1]{V0#1}

\newcommand\eqsp{\space}
\newcommand\yCM{{\ensuremath y_{\mathrm{CM}}}}
\newcommand\fTotal{{\ensuremath f_{\mathrm{total}}}}
\begin{document}
%
%
\begin{titlepage}
  \PHyear{2022}      
  \PHnumber{055}     
  \PHdate{17 March}  
  \title{System-size dependence of the charged-particle pseudorapidity
    density at $\FiveTeVBF$ for pp,  p\==Pb, and Pb\==Pb collisions}
  \ifplb
  \author{ALICE Collaboration}
  \fi
  \ShortTitle{System-size dependence of $\textdndeta$ at $\FiveTeV$}
  \Collaboration{ALICE Collaboration\thanks{See 
      Appendix~\ref{app:collab} for the list of collaboration members}}%
  \ShortAuthor{ALICE Collaboration} 

  \begin{abstract}
    We present the first systematic comparison of the charged-particle
    pseudorapidity densities for three widely different collision
    systems, pp, p\==Pb, and Pb\==Pb, at the top energy of the Large
    Hadron Collider (${\FiveTeV}$) measured over a wide pseudorapidity
    range (${-3.5 <\eta <5}$), the widest possible among the four
    experiments at that facility. The systematic uncertainties are
    minimised since the measurements are recorded by the same
    experimental apparatus (ALICE).  The distributions for p\==Pb and
    Pb\==Pb collisions are determined as a function of the centrality
    of the collisions, while results from pp collisions are reported
    for inelastic events with at least one charged particle at
    midrapidity.  The charged-particle pseudorapidity densities are,
    under simple and robust assumptions, transformed to
    charged-particle rapidity densities.  This allows for the
    calculation and the presentation of the evolution of the width of
    the rapidity distributions and of a lower bound on the Bjorken
    energy density, as a function of the number of participants in all
    three collision systems.  We find a decreasing width of the
    particle production, and roughly a smooth ten fold increase in the
    energy density, as the system size grows, which is consistent with
    a gradually higher dense phase of matter.
  \end{abstract}
\end{titlepage}
\setcounter{page}{2} 

\section{Introduction}

The number of charged particles produced in energetic nuclear
collisions is an important indicator for the strong interaction
processes that determine the particle production at the sub-nucleonic
level. In particular, the production of charged particles is expected
to reflect the number of quark and gluon collisions occurring during
the initial stages of the reaction. The total number of particles
produced also provides information on the energy transfer available
from the initial colliding beams to particle production, as a
consequence of nuclear stopping~\cite{Arsene:2009aa}.  In order to help
unravel this complex scenario it is important to compare the particle
production amongst collision systems of different sizes over a wide
kinematic range.

We present the measured charged-particle pseudorapidity density,
${\textdndeta}$, for pp, p\==Pb, and Pb\==Pb (previously
published~\cite{Adam:2016ddh}) collisions at the same collision energy
of ${\FiveTeV}$ in the nucleon--nucleon centre-of-mass reference
frame.  This is, at present, the maximum available energy at CERN's
Large Hadron Collider (LHC) for Pb\==Pb collisions. The measurements
were carried out using ALICE at LHC (for earlier $\textdndeta$ results
see for example
Refs.~\cite{Abreu:2002fw,PhysRevC.83.024913,ATLAS:2015hkr}). The three
studied reactions have different characteristics probing widely
different particle production yields and mechanisms. In Pb\==Pb
collisions, the total particle yield for central collisions is of the
order $10^4$~\cite{Adam:2016ddh}, and a strongly coupled plasma of
quarks and gluons (\sQgp{}) is formed
\cite{Arsene:2004fa,Back:2004je,Adams:2005dq,Adcox:2004mh}, whose
collective and transport properties are currently under intense
study. On the other hand, pp collisions represent the simplest
possible nuclear collision system, where the average total particle
production is much smaller (${\approx80}$, by integrating the measured
distributions), and is to first approximation much less subject to
collective effects~\cite{Bierlich:2021poz}. The p\==Pb system is
intermediate to the other reactions, corresponding to the situation
where a single nucleon probes the nucleons in a narrow cylinder of the
target nucleus. The extent to which p\==Pb is governed by the initial
state cold nuclear matter of the lead ion or whether collective
phenomena in the hot and dense medium play an important role is, at
present, a matter under scrutiny by the
community~\cite{Bierlich:2021poz,Lin:2021mdn}.

In this letter, we compare the three reactions and present the ratios
of the charged-particle pseudorapidity density distributions
(${\textdndeta}$) of the more complex reactions to the pp
distribution. Owing to ALICE's unique large acceptance in
pseudorapidity, and using simple and robust assumptions, we transform
the measured charged-particle pseudorapidity density distributions
into charged-particle rapidity density distributions
(${\textdndy}$). This allows us to calculate the width of the rapidity
distributions as a function of the number of participating nucleons.
The parameters of the transformation also allow us to estimate a lower
bound on the energy density using the well-known formula from
Bjorken~\cite{PhysRevD.27.140}.  An energy density exceeding the
critical energy density of roughly
$\unit[1]{\GeVfm}$~\cite{Ding:2014kva} is a necessary condition for
the formation of deconfined matter of quarks and gluons, and thus it
is of the utmost interest to understand the development of these
energy densities across different collision systems.

\section{Experimental set-up, data sample, analysis method, systematic 
  uncertainties}

A detailed description of the ALICE detector and its performance can
be found elsewhere~\cite{Aamodt:2008zz,Abelev:2014ffa}. The present
analysis uses the Silicon Pixel Detector (SPD) to determine the
pseudorapidity densities in the range ${-2 < \eta < 2}$ and the
Forward Multiplicity Detector (FMD) in the ranges
${-3.5 < \eta < -1.8}$ and ${1.8 < \eta < 5}$. The V0, comprised of
two plastic scintillator discs covering $-3.7<\eta<-1.7$ (V0C) and
$2.8<\eta<5.1$ (V0A), and the ZDC, two zero-degree calorimeters
located $\unit[112.5]{m}$ from the interaction point, measurements
determine the collision centrality and are used for offline event
selection~\cite{Adam:2016ddh}.

The results presented are based on data from collisions at a
centre-of-mass energy per nucleon pair of ${\FiveTeV}$ as collected by
ALICE during LHC Run~1~(2013) for p\==Pb, and during Run~2~(2015) for
pp and Pb\==Pb.  The FMD suffered high levels of background noise
during the 2016 p\==Pb campaign, due to the high collision rate, and
this data is therefore not used for the present analysis. About
${10^5}$ events with a minimum bias trigger
requirement~\cite{Adam:2016ddh} were analysed in the centrality range
from ${0\%}$ to ${90\%}$ and ${0\%}$ to ${100\%}$ of the visible cross
section for Pb\==Pb and p\==Pb collisions, respectively.  The minimum
bias trigger for p\==Pb and Pb\==Pb collisions in ALICE was defined as
a coincidence between the \VZERO{A} and \VZERO{C} sides of the V0
detector.

The data from the p\==Pb collisions were taken in two beam
configurations: one where the lead ion travelled toward positive
pseudorapidity and one where it travelled toward negative
pseudorapidity.  The results from the latter collisions are mirrored
around $\eta=0$.  The centre-of-mass frame in p\==Pb collisions does
not coincide with the laboratory frame, due to the single magnetic
field in the LHC, and thus the rapidity of the centre-of-mass is
$\yCM=\pm0.465$ for the two directions, respectively, in
the laboratory frame.  For this reason, pseudorapidity, calculated
with respect to the laboratory frame, is denoted ${\etalab}$ whenever
p\==Pb results are presented.

Likewise, for the pp collisions, about ${10^5}$ events with
coincidence between \VZERO{A} and \VZERO{C} and at least one charged
particle in ${|\eta|<1}$ were analysed.  By requiring at least one
charged particle at midrapidity, the so-called \INELGtZero{} event
class, the systematic uncertainty, related to the absolute
normalisation to the full inelastic cross section, is reduced, while
still sampling a large fraction (${>75\%}$) of the hadronic cross
section~\cite{Adam:2015gka,ALICE:2020swj}.

The standard ALICE event selection~\cite{Aamodt:2010pb} and centrality
estimator based on the \VZERO{}
amplitude~\cite{Abelev:2013qoq,Adam:2014qja} are used in this
analysis.  The event selection consists of: a) exclusion of background
events using the timing information from the ZDC (for Pb\==Pb and
p\==Pb, e.g., beam--gas interactions) and V0 detectors, b)
verification of the trigger conditions, and c) a reconstructed
position of the collision (primary vertex).  In Pb\==Pb collisions,
centrality is obtained from the sum amplitude in both V0 detector
arrays (V0M). For p\==Pb only the amplitude in the array on the
lead-going side (V0A or V0C) is used. In Pb\==Pb collisions, the 10\%
most peripheral collisions have substantial contributions from
electromagnetic processes and are therefore not included in the
results presented here~\cite{Abelev:2013qoq}.

A primary charged particle is defined as a charged particle with a
mean proper lifetime ${\tau}$ larger than
${\unit[1]{cm\kern-.05em/c}}$, which is either a) produced directly in
the interaction, or b) from decays of particles with ${\tau}$ smaller
than ${\unit[1]{cm\kern-.05em/c}}$~\cite{ALICE-PUBLIC-2017-005}. All
quantities reported here are for primary, charged particles, though
``primary'' is omitted in the following for brevity.

The analysis method is identical to that of previous
publications~\cite{Adam:2016ddh}: the measurement of the
charged-particle pseudorapidity density at midrapidity is obtained
from counting particle trajectories determined using the two layers of
the SPD. The SPD has a lower transverse momentum acceptance of
$\unit[50]{\MeVc{}}$, and the yield is extrapolated down to
${\pT=\unit[0]{\MeVc{}}}$ via simulations.  In the forward regions,
the measurement is provided by the analysis of the deposited energy
signal in the FMD and a statistical method is employed to calculate
the inclusive number of charged particles. A data-driven
correction~\cite{Adam:2015kda}, based on separate measurements
exploiting displaced collision vertices, is applied to remove the
background from secondary particles.

Systematic uncertainty estimations for the midrapidity measurements
are detailed elsewhere~\cite{Adam:2016ddh,Adam:2015gka,Adam:2014qja},
and are from background suppression, transverse momentum
extrapolation, weak decays, and simulations.  The estimates are
obtained through variation of thresholds and simulation studies. For
pp (p\==Pb), the total systematic uncertainty amounts to ${1.5\%}$
(${2.7\%}$) over the whole pseudorapidity range; while for Pb\==Pb the
total systematic uncertainty is ${2.6\%}$ at ${\eta=0}$ and ${2.9\%}$
at ${|\eta|=2}$. The systematic uncertainty is mostly correlated over
pseudorapidity for ${|\eta|<2}$, and largely independent of
centrality. The uncertainty in the forward region, estimated via
variations of thresholds and simulation studies, is the same for all
collision systems and is uncorrelated across ${\eta}$, amounting to
${6.9\%}$ for ${\eta>3.5}$ and ${6.4\%}$ elsewhere within the forward
regions~\cite{Adam:2015kda}.  In the figures of this letter,
uncorrelated, local in pseudorapidity, systematic uncertainties are
indicated by open boxes on the data points, while correlated
systematic uncertainties, those that affect the overall scale and
typically from event classification and selection, are indicated by
filled boxes to the right of the data.
The systematic uncertainty on ${\textdndeta}$, due to the centrality
class definition in Pb\==Pb, is estimated to vary from ${0.6\%}$ for
the most central to ${9.5\%}$ for the most peripheral
class~\cite{Adam:2015ptt}. The $80\%$ to $90\%$ centrality class has
residual contamination from electromagnetic processes as detailed
elsewhere~\cite{Abelev:2013qoq}, which gives rise to an additional
${4\%}$ systematic uncertainty in the measurements. No overall
systematic uncertainty has been estimated for p\==Pb collisions, as
the centrality selection in that collision system is inherently
difficult to map to the underlying dynamics of the
collisions~\cite{Adam:2014qja}.

\section{Results}

\begin{figure}[t!]
  \centering
  \begin{tabular}{@{}p{.8\linewidth}@{}}
    \includegraphics[width=\linewidth]{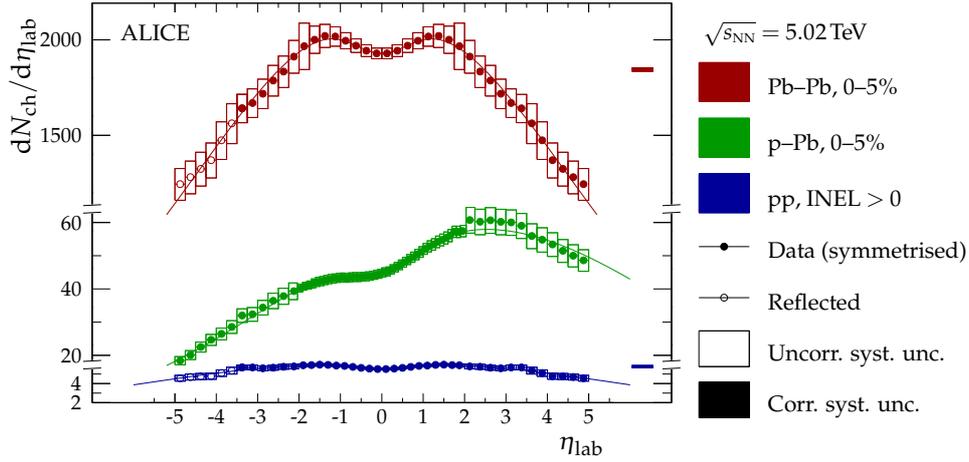}
  \end{tabular}
  \caption[${\textdndetalab}$ at ${\FiveTeV{}}$ in Pb\==Pb, p\==Pb
  (most central), and for pp collisions with \INELGtZero{}
  trigger.]{Charged-particle pseudorapidity density in
    Pb\==Pb~\cite{Adam:2016ddh} and p\==Pb for the $5\%$ most central
    collisions, and for pp collisions with \INELGtZero{} trigger
    class.  For symmetric collision systems (Pb\==Pb and pp) the data
    has been symmetrised around ${\eta=0}$ and points for ${\eta>3.5}$
    have been reflected around ${\eta=0}$.  The boxes around the points
    and to the right reflect the uncorrelated and correlated, with
    respect to pseudorapidity, systematic uncertainty, respectively.
    The relative correlated, normalisation, uncertainties are
    evaluated at $\textdndeta|_{\eta=0}$.  The lines
    show fits of \EqRef{eq:f} (Pb\==Pb and pp) and \EqRef{eq:g}
    (p\==Pb) to the data (discussed in
    Section~\ref{sec:discussion}). Please note that the ordinate has
    been cut twice to accommodate for the very different ranges of the
    charged-particle pseudorapidity densities.}
  \label{fig:all:central}   
\end{figure}

\figureref{fig:all:central} shows the measured pseudorapidity
densities in pp, and in central p\==Pb, and the previously published
results for Pb\==Pb~\cite{Adam:2016ddh} collisions at ${\FiveTeV}$ for
primary particles.

For the ${5\%}$ most central Pb\==Pb collisions
${\textdndeta \approx 2000 }$ at midrapidity
(${\eta=0}$)~\cite{Adam:2016ddh}, while for p\==Pb collisions the
distribution peaks at ${\textdndetalab \approx 60 }$ around ${\etalab=3}$
in the lead-going direction (${\eta>0}$).  For pp collisions with the
\INELGtZero{} trigger condition discussed above,
${\textdndeta=5.7\pm0.2}$ at midrapidity, consistent with previous
results derived from ${\pT}$ spectra~\cite{ALICE:2019dfi}.

\begin{figure}[t!]
  \centering
  \begin{tabular}{@{}p{.8\linewidth}@{}}
    \includegraphics[width=\linewidth]{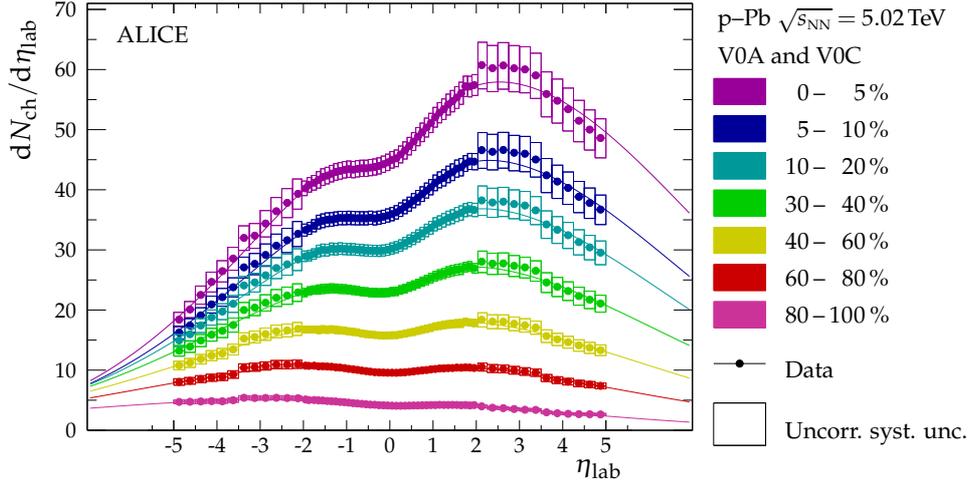}
  \end{tabular}
  \caption[${\textdndetalab}$ in p\==Pb collisions at
  ${\FiveTeV}${}.]{ Charged-particle pseudorapidity density in p\==Pb
    collisions at ${\FiveTeV}$ in seven centrality classes based on
    the V0A and V0C estimators.  The lines are obtained using a fit of
    a scaled, normal distribution in rapidity Eq.~\eqref{eq:g} to the
    data (discussed in Section~\ref{sec:discussion}).}
  \label{fig:pPb:cent}
\end{figure}

\figureref{fig:pPb:cent} shows, as a function of centrality, the
measured charged-particle pseudorapidity densities for p\==Pb
collisions at ${\FiveTeV}$. The strategy of centrality selection for
proton on nucleus reactions is explained
elsewhere~\cite{Adam:2014qja}.  The ALICE Collaboration has previously
presented $\textdndeta$ for Pb\==Pb collisions at this
energy~\cite{Adam:2016ddh}.

\begin{figure}[t!]
  \centering
  \begin{tabular}{@{}p{.8\linewidth}@{}}
    \includegraphics[width=\linewidth]{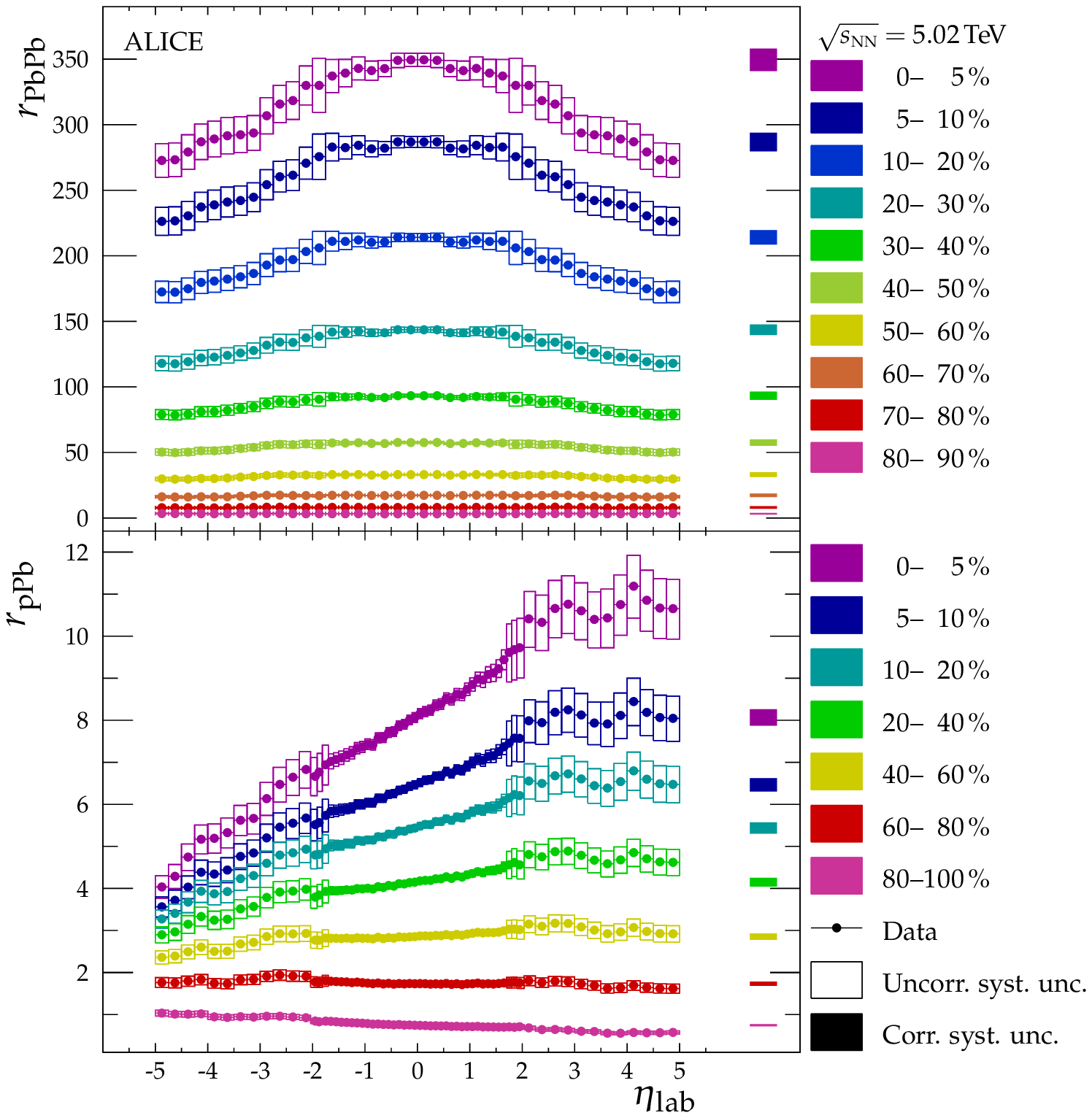}
  \end{tabular}

  \caption[Ratio of ${\textdndetalab}$ in p\==Pb and Pb\==Pb to pp]{%
    Ratio ${r_X}$ of the charged-particle pseudorapidity density in
    Pb\==Pb (top) and p\==Pb (bottom) in different centrality classes
    to the charged-particle pseudorapidity density in pp in the
    \INELGtZero{} event class. Note, for Pb\==Pb ${\etalab}$ is the
    same as the centre-of-mass pseudorapidity.}
  \label{fig:all:rX}
\end{figure}    

In \figref{fig:all:rX}, the charged-particle pseudorapidity densities
in p\==Pb and Pb\==Pb reactions are divided by the pp distributions
corresponding to the \INELGtZero{} trigger class. The ratio is
${r_{X}={{(\textdndeta|_{X}})/({\textdndeta|_{\mathrm{pp}}})}}$, where
${X}$ labels p\==Pb and Pb\==Pb collisions, in centrality classes, as
a function of pseudorapidity.  In the ratios, systematic
uncertainties, of common origin, are partially cancelled, and, as an
estimate, the magnitude of the resulting systematic uncertainties are
given only by the uncertainties in the ${\textdndeta|_{X}}$
measurements, since the uncertainties are independent of the collision
system. In p\==Pb collisions the rapidity of the centre-of-mass is
non-zero, which is not taken into account in the ratios. Such a
correction would require prior determination of the full Jacobian of
the transformation from pseudorapidity to rapidity, which is not
possible to perform reliably with the ALICE apparatus.

The ratio of the p\==Pb relative to the pp distributions increases
with pseudorapidity from the p-going to the Pb-going direction for
central collisions, which Brodsky~\emph{et~al} and
Adil~\emph{et~al}~\cite{Brodsky:1977de,Adil:2005qn} suggest is a sign
of scaling of the pp distribution with the increasing number of
participants as the lead nucleus is probed by the incident proton, and
thus independent proton--nucleon scatterings on the lead-ion side. A
similar scaling, however, does not hold for the Pb\==Pb reaction.  The
ratios cannot be obtained by simple scaling of the elementary pp
distributions.  Instead, the ratio of the Pb\==Pb relative to the pp
distributions exhibits an enhancement of particle production around
midrapidity for the more central collisions which is indicative of the
formation of the \sQgp{}~\cite{Back:2004je}.  Likewise,
${r_{\textrm{pPb}}}$ increases for all but the two most peripheral
centrality classes as ${\etalab\rightarrow 3}$.  In Pb\==Pb collisions
it is seen that the various mechanisms behind the pseudorapidity
distributions are more transversely directed than in pp collisions by
the increase of ${r_{\textrm{PbPb}}}$ as ${|\eta|\rightarrow0}$

\section{Rapidity and energy-density dependence on system size and
  discussion }
\label{sec:discussion}

It has been shown that the charged-particle \textit{rapidity} density
(${\textdndy}$) in Pb\==Pb collisions, to a good accuracy, follows a
normal distribution over the considered rapidity interval
($|y|\lesssim5$)~\cite{Adam:2016ddh,Abbas:2013bpa}.  Those results
relied on calculating the average Jacobian
${\textdndy=\langle J\rangle=\langle \beta\rangle}$
using the full ${\pT}$ spectra, at midrapidity, of charged
pions and kaons as well as protons and antiprotons.  Here, we use the
approximation
\begin{linenomath*}
  \begin{align*}
    y\approx\eta-\frac12\frac{m^2}{\pT^2}\cos\vartheta\eqsp,
  \end{align*}
\end{linenomath*}
where ${\vartheta}$ is the polar angle of emission, and identify
${a=\pT/m}$ with an effective ratio of transverse momentum
over mass. With this, the effective Jacobian can be written as
\begin{linenomath*}
  \begin{align*}
    J'(\eta,a)
    &=
      {\textstyle \left(1+\frac{1}{a^2}\frac{1}{\cosh^2\eta}\right)^{-1/2}}\eqsp.
  \end{align*}
\end{linenomath*}

We further make the ansatz that ${\textdndy}$ is normal distributed
for symmetric collision systems (pp and Pb\==Pb), so that
${\textdndeta}$ can be parameterised as
\begin{linenomath*}
  \begin{align}
    \label{eq:f}
    f(\eta;A,a,\sigma)
    &=
      J'(\eta,a)\,A{\textstyle\frac{1}{\sqrt{2\pi}\sigma}
      \exp\left({-\frac{y^2\{\eta,a\}}{2\sigma^2}}\right)}\eqsp, 
  \end{align}
\end{linenomath*}
where ${A}$ and $\sigma$ are the total integral and
width of the distribution, respectively, and ${y}$ the rapidity in the
centre-of-mass frame.  Motivated by the observed approximate linearity
of ${r_{\mathrm{pPb}}}$ (see lower panel of \figref{fig:all:rX}), we
replace ${A}$ with ${(\alpha y + A)}$ for the asymmetric system
(p\==Pb) and parameterise ${\textdndetalab}$ as
\begin{linenomath*}
  \begin{align}
    \label{eq:g}
    g(\eta;A,a,\alpha,\sigma)
    &=
      J'(\eta,a)\,\left(\alpha y\{\eta,a\}+A\right)
      {\textstyle\frac{1}{\sqrt{2\pi}\sigma}
      \exp\left({-\frac{\left[y\{\eta,a\}-\yCM\right]^2}{
      2\sigma^2}}\right)}
      \eqsp.
  \end{align}
\end{linenomath*}

\begin{figure}[ht!]
  \centering
  \begin{tabular}{@{}p{.8\linewidth}@{}}
    \includegraphics[width=\linewidth]{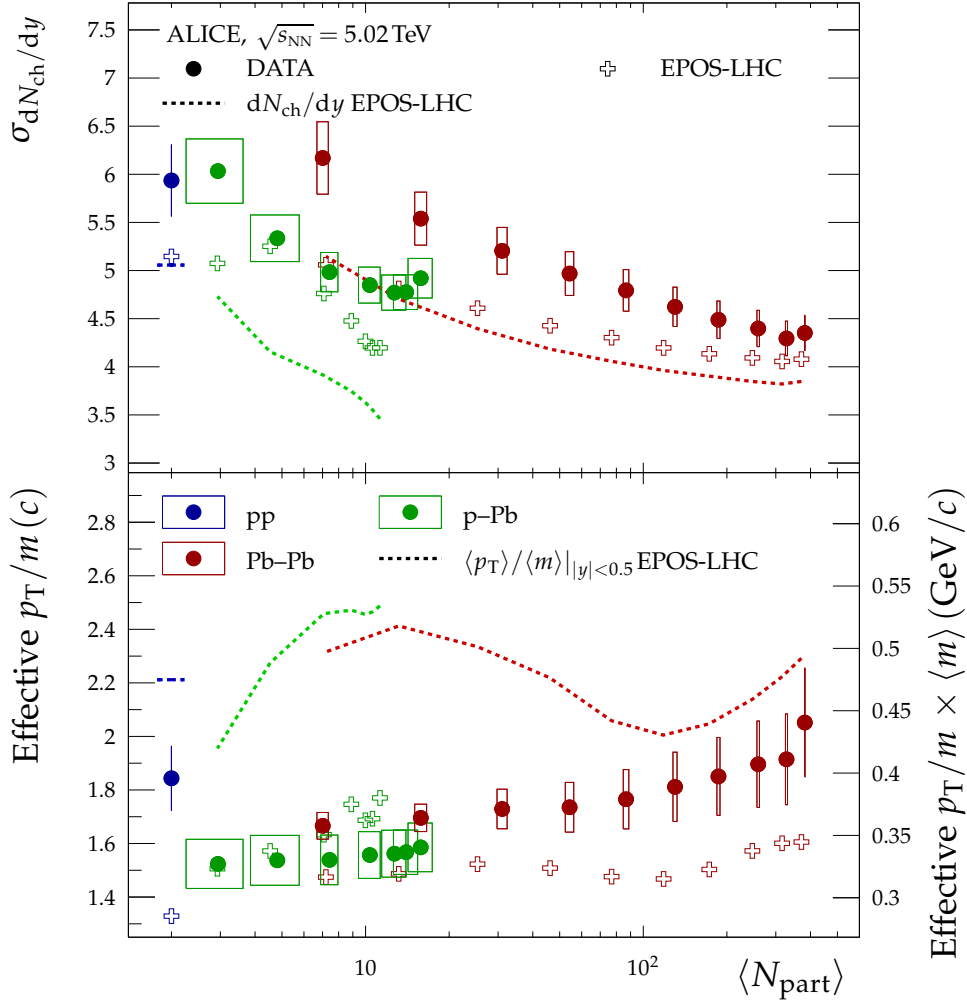}
    \\
  \end{tabular}
  \caption[Fit parameters of scaled normal distribution to pp, p\==Pb,
  and Pb\==Pb ${\textdndy}$]{%
    The width (top) and effective ${\textpTm}$ (bottom) fit parameters
    as a function of the mean number of participants in pp, p\==Pb,
    and Pb\==Pb collisions at ${\FiveTeV}$.
    Vertical uncertainties are the standard error on the best-fit
    parameter values, while horizontal uncertainties reflect the
    uncertainty on $\langle\Npart\rangle$ from the Glauber
    calculations. 
    Also shown are similar fit parameters from the same
    parameterisation of EPOS-LHC calculations as well as direct
    calculations of the standard deviation of the $\textdndy$
    distributions and the $\textavgpTm$ ratio from the EPOS-LHC
    calculations.}
  \label{fig:all:pars}
\end{figure}   

The functions ${f}$ and ${g}$ defined in \EqRef{eq:f} and
\EqRef{eq:g}, respectively, describe the measurements within the
measured region with $\chi^2$ per degrees of freedom ($\nu$) in the
range of $0.1$ to $0.5$.  The small $\chi^2/\nu$ values are a
consequence of the relatively large uncorrelated systematic
uncertainties on the measurements.  That is, the charged-particle
distributions for pp, p\==Pb, and Pb\==Pb collisions at ${\FiveTeV}$
follow a normal distribution in rapidity, with free parameters $A$, $a$,
$\sigma$, and $\alpha$ in the asymmetric case.

The top panel of \figref{fig:all:pars} shows the best-fit parameter
values of the normal width (${\sigma_{\textdndy}}$) for all three
collision systems as a function of the average number of participating
nucleons (${\langle\Npart\rangle}$) calculated using a Glauber
model~\cite{ALICE-PUBLIC-2018-011}.  The best-fit parameters are found
taking statistical and uncorrelated systematic uncertainties into
account. 
The result using the above procedure, for the most central Pb\==Pb
collisions, is found to be compatible with previous results extracted
by unfolding with the mean Jacobian estimated from transverse momentum
spectra~\cite{Adam:2016ddh}. %
The open points (crosses) and dashed lines on the figure are from
evaluations of \EqRef{eq:f} and \EqRef{eq:g}, and direct calculations
of ${\sigma_{\textdndy}}$, respectively, using model calculations with
EPOS-LHC~\cite{Pierog:2013ria}. EPOS-LHC was chosen as it provides
predictions for all three collision systems.  The parameterisation, in
terms of the two functions, of this model calculation generally
reproduces the widths of the charged-particle rapidity densities,
except in the asymmetric case where a direct evaluation of the
standard deviation is less motivated.

The general trend is that the widths decrease as
${\langle\Npart\rangle}$ increases, consistent with the
behaviour of the ${r_{\textrm{PbPb}}}$ ratios.  Notably, the width
of the ${\textdndy}$ distributions in p\==Pb and Pb\==Pb, for low
number of participant nucleons in the collisions, approaches the width
of the pp distribution, which, presumably, is dominated by kinematic
and phase space constraints.

The lower panel of \figref{fig:all:pars} shows the dependence of ${a}$
on the average number of participants.  The right-hand ordinate is the
same, but multiplied by the average mass
${\langle m\rangle=\unit[\left(0.215\pm0.001\right)]{\GeVc}}$ estimated
from measurements of identified particles in Pb\==Pb collisions at
$\TwoTeV$~\cite{Abelev:2013vea}. 
To better understand the parameter $a$, this parameter extracted from
the EPOS-LHC calculations, using the above procedure, is also shown in
the figure.  The dotted lines show the average ${\pT/m}$ predicted by
EPOS-LHC~\cite{Pierog:2013ria}.  The EPOS-LHC calculations indicate
that the extracted effective transverse momentum to mass ratio $a$ is
consistently smaller than the ratio of the average transverse momentum
to the average mass.  Thus $a$ gives a lower bound on ${\textavgpTm}$.

We can estimate the energy density that is reached in the collisions
as a function of the number of participants for the three systems. A
conventional approach is to use the model originally proposed by
Bjorken~\cite{PhysRevD.27.140} in which the energy density (${\Ebj}$)
depends on the rapidity density of particles and the volume of a
longitudinal cylinder with cross sectional area determined by the
overlap between the colliding partners and length determined by a
characteristic particle formation time
\begin{linenomath*}
  \begin{align*}
    \Ebj &= \frac{1}{c\tau}\frac1{\ST}\left\langle\dispdEdy\right\rangle\eqsp.
  \end{align*}
\end{linenomath*}
Here, ${\ST\approx \pi R^2\approx\pi\Npart^{2/3}}$ is the transverse area
spanned by the participating nucleons, ${\textdEdy}$ is the
transverse-energy rapidity density, and ${\tau}$ is the formation time.
While a formation time of ${\tau=\unit[1]{\mathrm{fm}/\kern-.05emc}}$ is
often assumed, it is left as a free parameter here.  With
${\langle\mT\rangle=\langle m\rangle\sqrt{1+(\textavgpTm)^2}}$, the
transverse-energy rapidity density can be approximated by
\begin{linenomath*}
  \begin{align*}
    \left\langle\dispdEdy\right\rangle
    &\approx \langle\mT\rangle \frac1{\fTotal}\dispdndy
      = \langle m\rangle\sqrt{1+\left(\dispavgpTm\right)^2}\frac1{\fTotal}
      \dispdndy
      \eqsp,
  \end{align*}
\end{linenomath*}
where ${\fTotal=0.55\pm0.01}$, the ratio of charged particles to all
particles~\cite{Adam:2016thv}, accounts for neutral particles not
measured in the experiment, and is assumed the same for all collision
systems.  Substituting the derived ${\textdndy}$ and the effective
${a=\textpTm\lesssim\textavgpTm}$ results in a lower bound estimate
for the Bjorken energy density (${\Elb}$)
\begin{linenomath*}
  \begin{align}
    \Ebj\tau &\ge \Elb\tau
             = \frac1c\frac1{\ST}\langle
               m\rangle\sqrt{1+a^2}\frac1{\fTotal}
               \sqrt{1+\frac1{a^2}\frac1{\cosh^2\eta}}
               \dispdndeta
               \eqsp,
               \label{eq:elb}
  \end{align}
\end{linenomath*}
where $a$ and $\langle m\rangle$ are as in the top panel of
\figref{fig:all:pars}.

\begin{figure}[t!]
  \centering
  \begin{tabular}{@{}p{.6\linewidth}@{}}
    \includegraphics[width=\linewidth]{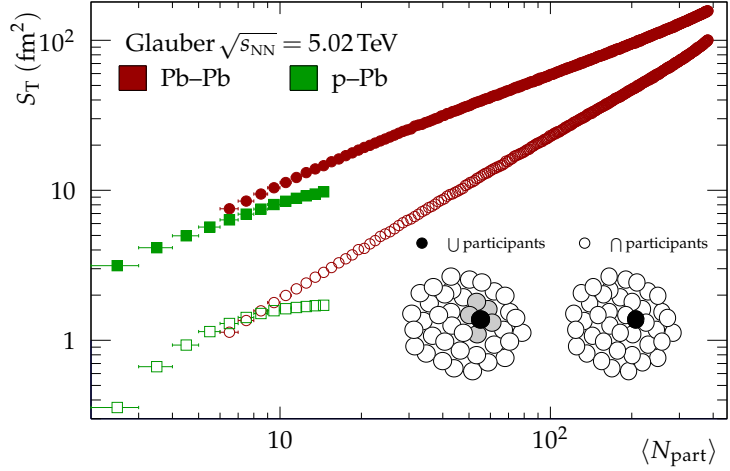}
  \end{tabular}
  \caption[Transverse area Glauber calculation]{%
    The transverse area ${\ST}$ as calculated in a numerical Glauber
    model for two extreme cases: a) only the exclusive overlap of
    nucleons is considered (${\cap}$, open markers) and b) the
    inclusive area of participating nucleons contribute (${\cup}$,
    closed markers) in both p\==Pb and Pb\==Pb at ${\FiveTeV}$.}
  \label{fig:all:area}
\end{figure}

The transverse area ${\ST}$ is estimated in a numerical Glauber
model~\cite{LOIZIDES201513,Loizides:2016djv} as shown in
\figref{fig:all:area}.  We consider two extremes for the transverse
area spanned by the participating nucleons: a) the \emph{exclusive}
(or direct) overlap between participating nucleons, ${\cap}$ and open
markers in \figref{fig:all:area}, and b) the \emph{inclusive} (or
full) area of all participating nucleons, ${\cup}$ and full markers in
\figref{fig:all:area}.

\begin{figure}[ht!]
  \centering
  \begin{tabular}{@{}p{.8\linewidth}@{}}
    \includegraphics[width=\linewidth]{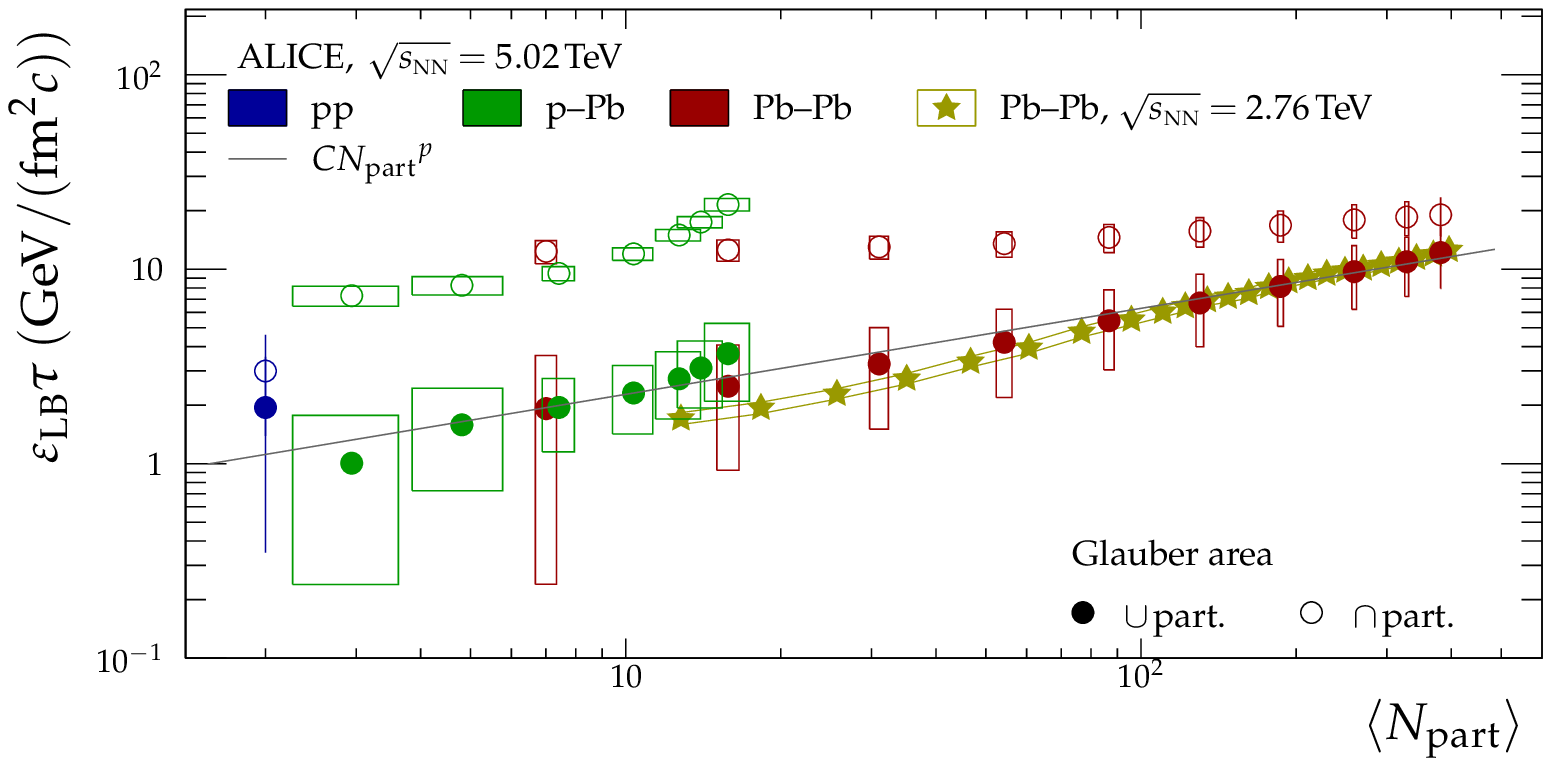}\\
  \end{tabular}
  \caption[Lower bound estimate of the transverse energy density]{%
    Estimate of the lower bound on the Bjorken transverse energy
    density in pp, p\==Pb, and Pb\==Pb collisions at ${\FiveTeV}$,
    considering the exclusive ($\cap$, open markers) and inclusive
    ($\cup$, full markers) overlap area ${\ST}$ of the nucleons.  The
    expression ${C\Npart{}^p}$ is fitted to case ${\cup}$, and we find
    ${C=\unit[(0.8\pm0.3)]{\GeVfmc}}$ and ${p=0.44\pm0.08}$.
    Also shown is an estimate, via $\textdEdy$, of $\Ebj$ from
    Pb\==Pb collisions at ${\TwoTeV}$ (stars with uncertainty
    band)~\cite{Adam:2016thv}.}
  \label{fig:all:elb}
\end{figure}

\figureref{fig:all:elb} shows the lower-bound energy density estimate,
${\Elb\tau\le\Ebj\tau}$, as a function of the number of participants,
which reaches values between ${10}$ and ${\unit[20]{\GeVfmc}}$ in the
most central Pb\==Pb collisions. The uncertainties are from standard
error propagation of Eq.~\eqref{eq:elb} of uncertainties on the
best-fit parameter values, the number of participants, mean mass, and
${\fTotal}$. A rise from roughly ${\unit[1]{\GeVfmc}}$ to over
${\unit[10]{\GeVfmc}}$ is observed if the transverse area is assumed
to be the inclusive area of participating nucleons.  This trend is
illustrated by a power-law (${C\Npart^p}$) fit to the data in the
figure, with the parameter values ${C=\unit[(0.8\pm0.3)]{\GeVfmc}}$
and ${p=0.44\pm0.08}$.  On the other hand, if the transverse area is
assumed to be the smaller exclusive overlap area, we observe a
substantially larger lower bound on the energy density, but a less
dramatic increase with increasing number of participating nucleons.
Also shown in the figure are estimates of the Bjorken energy density
${\Ebj\tau}$ for Pb\==Pb reactions at
${\TwoTeV}$~\cite{Adam:2016thv}. These results where obtained from
measurements of the transverse energy in the collisions and using the
inclusive estimate of the transverse area $\ST$. 
The trend of the ${\FiveTeV}$ results are similar to these earlier
results.  Bearing in mind that for the largest LHC collision energy we
show a lower bound estimate of the energy density in
\figref{fig:all:elb}, we find a likely overall increase in the energy
density from ${\TwoTeV}$ to ${\unit[5.02]{\TeV}}$.

\section{Summary and conclusions}

We have measured the charged particle pseudorapidity density in pp,
p\==Pb, and Pb\==Pb collisions at ${\FiveTeV}$ over the widest
possible pseudorapidity range available at the LHC.  The distributions
where determined using the same experimental apparatus and methods,
and systematic uncertainties have been minimised to within the
capabilities of the set-up.  While the particle production in central
Pb\==Pb collisions clearly exhibits an enhancement as compared to pp
collisions, particle production in p\==Pb collisions is consistent
with dominantly incoherent nucleon--nucleon collisions.  By
transforming the measured pseudorapidity distributions to rapidity
distributions we have obtained systematic trends for the width of the
rapidity distributions and a lower bound on the energy density, which
shows a clear scaling behaviour as a function of the average number of
participant nucleons. The decreasing width of the deduced rapidity
distributions with increasing participant number suggests that the
kinematic spread of particles, including longitudinal degrees of
freedom, is reduced due to interactions in the early stages of the
collisions.  This is also reflected in the accompanying growth of the
energy density. Both observations are consistent with the gradual
establishment of a high-density phase of matter with increasing size
of the collision domain.

\newenvironment{acknowledgement}{\relax}{\relax}
\begin{acknowledgement}
\section*{Acknowledgements}

The ALICE Collaboration would like to thank all its engineers and technicians for their invaluable contributions to the construction of the experiment and the CERN accelerator teams for the outstanding performance of the LHC complex.
The ALICE Collaboration gratefully acknowledges the resources and support provided by all Grid centres and the Worldwide LHC Computing Grid (WLCG) collaboration.
The ALICE Collaboration acknowledges the following funding agencies for their support in building and running the ALICE detector:
A. I. Alikhanyan National Science Laboratory (Yerevan Physics Institute) Foundation (ANSL), State Committee of Science and World Federation of Scientists (WFS), Armenia;
Austrian Academy of Sciences, Austrian Science Fund (FWF): [M 2467-N36] and Nationalstiftung f\"{u}r Forschung, Technologie und Entwicklung, Austria;
Ministry of Communications and High Technologies, National Nuclear Research Center, Azerbaijan;
Conselho Nacional de Desenvolvimento Cient\'{\i}fico e Tecnol\'{o}gico (CNPq), Financiadora de Estudos e Projetos (Finep), Funda\c{c}\~{a}o de Amparo \`{a} Pesquisa do Estado de S\~{a}o Paulo (FAPESP) and Universidade Federal do Rio Grande do Sul (UFRGS), Brazil;
Ministry of Education of China (MOEC) , Ministry of Science \& Technology of China (MSTC) and National Natural Science Foundation of China (NSFC), China;
Ministry of Science and Education and Croatian Science Foundation, Croatia;
Centro de Aplicaciones Tecnol\'{o}gicas y Desarrollo Nuclear (CEADEN), Cubaenerg\'{\i}a, Cuba;
Ministry of Education, Youth and Sports of the Czech Republic, Czech Republic;
The Danish Council for Independent Research | Natural Sciences, the VILLUM FONDEN and Danish National Research Foundation (DNRF), Denmark;
Helsinki Institute of Physics (HIP), Finland;
Commissariat \`{a} l'Energie Atomique (CEA) and Institut National de Physique Nucl\'{e}aire et de Physique des Particules (IN2P3) and Centre National de la Recherche Scientifique (CNRS), France;
Bundesministerium f\"{u}r Bildung und Forschung (BMBF) and GSI Helmholtzzentrum f\"{u}r Schwerionenforschung GmbH, Germany;
General Secretariat for Research and Technology, Ministry of Education, Research and Religions, Greece;
National Research, Development and Innovation Office, Hungary;
Department of Atomic Energy Government of India (DAE), Department of Science and Technology, Government of India (DST), University Grants Commission, Government of India (UGC) and Council of Scientific and Industrial Research (CSIR), India;
National Research and Innovation Agency - BRIN, Indonesia;
Istituto Nazionale di Fisica Nucleare (INFN), Italy;
Japanese Ministry of Education, Culture, Sports, Science and Technology (MEXT) and Japan Society for the Promotion of Science (JSPS) KAKENHI, Japan;
Consejo Nacional de Ciencia (CONACYT) y Tecnolog\'{i}a, through Fondo de Cooperaci\'{o}n Internacional en Ciencia y Tecnolog\'{i}a (FONCICYT) and Direcci\'{o}n General de Asuntos del Personal Academico (DGAPA), Mexico;
Nederlandse Organisatie voor Wetenschappelijk Onderzoek (NWO), Netherlands;
The Research Council of Norway, Norway;
Commission on Science and Technology for Sustainable Development in the South (COMSATS), Pakistan;
Pontificia Universidad Cat\'{o}lica del Per\'{u}, Peru;
Ministry of Education and Science, National Science Centre and WUT ID-UB, Poland;
Korea Institute of Science and Technology Information and National Research Foundation of Korea (NRF), Republic of Korea;
Ministry of Education and Scientific Research, Institute of Atomic Physics, Ministry of Research and Innovation and Institute of Atomic Physics and University Politehnica of Bucharest, Romania;
Ministry of Education, Science, Research and Sport of the Slovak Republic, Slovakia;
National Research Foundation of South Africa, South Africa;
Swedish Research Council (VR) and Knut \& Alice Wallenberg Foundation (KAW), Sweden;
European Organization for Nuclear Research, Switzerland;
Suranaree University of Technology (SUT), National Science and Technology Development Agency (NSTDA), Thailand Science Research and Innovation (TSRI) and National Science, Research and Innovation Fund (NSRF), Thailand;
Turkish Energy, Nuclear and Mineral Research Agency (TENMAK), Turkey;
National Academy of  Sciences of Ukraine, Ukraine;
Science and Technology Facilities Council (STFC), United Kingdom;
National Science Foundation of the United States of America (NSF) and United States Department of Energy, Office of Nuclear Physics (DOE NP), United States of America.
In addition, individual groups or members have received support from:
Marie Sk\l{}odowska Curie, Strong 2020 - Horizon 2020 (grant nos. 824093, 896850), European Union;
Academy of Finland (Center of Excellence in Quark Matter) (grant nos. 346327, 346328), Finland.

\end{acknowledgement}

\ifplb
\bibliographystyle{elsarticle-num}
\else
\bibliographystyle{utphys}
\fi
\bibliography{bib}

\clearpage
\appendix
\section{The ALICE Collaboration}
\label{app:collab}
\begin{flushleft} 
\small

S.~Acharya\,\orcidlink{0000-0002-9213-5329}\,$^{\rm 123,130}$, 
D.~Adamov\'{a}\,\orcidlink{0000-0002-0504-7428}\,$^{\rm 85}$, 
A.~Adler$^{\rm 68}$, 
G.~Aglieri Rinella\,\orcidlink{0000-0002-9611-3696}\,$^{\rm 32}$, 
M.~Agnello\,\orcidlink{0000-0002-0760-5075}\,$^{\rm 29}$, 
N.~Agrawal\,\orcidlink{0000-0003-0348-9836}\,$^{\rm 49}$, 
Z.~Ahammed\,\orcidlink{0000-0001-5241-7412}\,$^{\rm 130}$, 
S.~Ahmad\,\orcidlink{0000-0003-0497-5705}\,$^{\rm 15}$, 
S.U.~Ahn\,\orcidlink{0000-0001-8847-489X}\,$^{\rm 69}$, 
I.~Ahuja\,\orcidlink{0000-0002-4417-1392}\,$^{\rm 36}$
% ...

\section*{Affiliation Notes}

$^{\rm I}$ Deceased\\
$^{\rm II}$ Also at: Max-Planck-Institut f\"{u}r Physik, Munich, Germany\\
$^{\rm III}$ Also at: Italian National Agency for New Technologies, Energy and Sustainable Economic Development (ENEA), Bologna, Italy\\
$^{\rm IV}$ Also at: Dipartimento DET del Politecnico di Torino, Turin, Italy\\
$^{\rm V}$ Also at: Department of Applied Physics, Aligarh Muslim University, Aligarh, India\\
$^{\rm VI}$ Also at: Institute of Theoretical Physics, University of Wroclaw, Poland\\
$^{\rm VII}$ Also at: An institution covered by a cooperation agreement with CERN\\

\section*{Collaboration Institutes}

$^{1}$ A.I. Alikhanyan National Science Laboratory (Yerevan Physics Institute) Foundation, Yerevan, Armenia\\
$^{2}$ AGH University of Science and Technology, Cracow, Poland\\
$^{3}$ Bogolyubov Institute for Theoretical Physics, National Academy of Sciences of Ukraine, Kiev, Ukraine\\
$^{4}$ Bose Institute, Department of Physics  and Centre for Astroparticle Physics and Space Science (CAPSS), Kolkata, India\\
$^{5}$ California Polytechnic State University, San Luis Obispo, California, United States\\
$^{6}$ Central China Normal University, Wuhan, China\\
$^{7}$ Centro de Aplicaciones Tecnol\'{o}gicas y Desarrollo Nuclear (CEADEN), Havana, Cuba\\
$^{8}$ Centro de Investigaci\'{o}n y de Estudios Avanzados (CINVESTAV), Mexico City and M\'{e}rida, Mexico\\
$^{9}$ Chicago State University, Chicago, Illinois, United States\\
$^{10}$ China Institute of Atomic Energy, Beijing, China\\
$^{11}$ Chungbuk National University, Cheongju, Republic of Korea\\
$^{12}$ Comenius University Bratislava, Faculty of Mathematics, Physics and Informatics, Bratislava, Slovak Republic\\
$^{13}$ COMSATS University Islamabad, Islamabad, Pakistan\\
$^{14}$ Creighton University, Omaha, Nebraska, United States\\
$^{15}$ Department of Physics, Aligarh Muslim University, Aligarh, India\\
$^{16}$ Department of Physics, Pusan National University, Pusan, Republic of Korea\\
$^{17}$ Department of Physics, Sejong University, Seoul, Republic of Korea\\
$^{18}$ Department of Physics, University of California, Berkeley, California, United States\\
$^{19}$ Department of Physics, University of Oslo, Oslo, Norway\\
$^{20}$ Department of Physics and Technology, University of Bergen, Bergen, Norway\\
$^{21}$ Dipartimento di Fisica, Universit\`{a} di Pavia, Pavia, Italy\\
$^{22}$ Dipartimento di Fisica dell'Universit\`{a} and Sezione INFN, Cagliari, Italy\\
$^{23}$ Dipartimento di Fisica dell'Universit\`{a} and Sezione INFN, Trieste, Italy\\
$^{24}$ Dipartimento di Fisica dell'Universit\`{a} and Sezione INFN, Turin, Italy\\
$^{25}$ Dipartimento di Fisica e Astronomia dell'Universit\`{a} and Sezione INFN, Bologna, Italy\\
$^{26}$ Dipartimento di Fisica e Astronomia dell'Universit\`{a} and Sezione INFN, Catania, Italy\\
$^{27}$ Dipartimento di Fisica e Astronomia dell'Universit\`{a} and Sezione INFN, Padova, Italy\\
$^{28}$ Dipartimento di Fisica `E.R.~Caianiello' dell'Universit\`{a} and Gruppo Collegato INFN, Salerno, Italy\\
$^{29}$ Dipartimento DISAT del Politecnico and Sezione INFN, Turin, Italy\\
$^{30}$ Dipartimento di Scienze MIFT, Universit\`{a} di Messina, Messina, Italy\\
$^{31}$ Dipartimento Interateneo di Fisica `M.~Merlin' and Sezione INFN, Bari, Italy\\
$^{32}$ European Organization for Nuclear Research (CERN), Geneva, Switzerland\\
$^{33}$ Faculty of Electrical Engineering, Mechanical Engineering and Naval Architecture, University of Split, Split, Croatia\\
$^{34}$ Faculty of Engineering and Science, Western Norway University of Applied Sciences, Bergen, Norway\\
$^{35}$ Faculty of Nuclear Sciences and Physical Engineering, Czech Technical University in Prague, Prague, Czech Republic\\
$^{36}$ Faculty of Science, P.J.~\v{S}af\'{a}rik University, Ko\v{s}ice, Slovak Republic\\
$^{37}$ Frankfurt Institute for Advanced Studies, Johann Wolfgang Goethe-Universit\"{a}t Frankfurt, Frankfurt, Germany\\
$^{38}$ Fudan University, Shanghai, China\\
$^{39}$ Gangneung-Wonju National University, Gangneung, Republic of Korea\\
$^{40}$ Gauhati University, Department of Physics, Guwahati, India\\
$^{41}$ Helmholtz-Institut f\"{u}r Strahlen- und Kernphysik, Rheinische Friedrich-Wilhelms-Universit\"{a}t Bonn, Bonn, Germany\\
$^{42}$ Helsinki Institute of Physics (HIP), Helsinki, Finland\\
$^{43}$ High Energy Physics Group,  Universidad Aut\'{o}noma de Puebla, Puebla, Mexico\\
$^{44}$ Horia Hulubei National Institute of Physics and Nuclear Engineering, Bucharest, Romania\\
$^{45}$ Indian Institute of Technology Bombay (IIT), Mumbai, India\\
$^{46}$ Indian Institute of Technology Indore, Indore, India\\
$^{47}$ INFN, Laboratori Nazionali di Frascati, Frascati, Italy\\
$^{48}$ INFN, Sezione di Bari, Bari, Italy\\
$^{49}$ INFN, Sezione di Bologna, Bologna, Italy\\
$^{50}$ INFN, Sezione di Cagliari, Cagliari, Italy\\
$^{51}$ INFN, Sezione di Catania, Catania, Italy\\
$^{52}$ INFN, Sezione di Padova, Padova, Italy\\
$^{53}$ INFN, Sezione di Pavia, Pavia, Italy\\
$^{54}$ INFN, Sezione di Torino, Turin, Italy\\
$^{55}$ INFN, Sezione di Trieste, Trieste, Italy\\
$^{56}$ Inha University, Incheon, Republic of Korea\\
$^{57}$ Institute for Gravitational and Subatomic Physics (GRASP), Utrecht University/Nikhef, Utrecht, Netherlands\\
$^{58}$ Institute of Experimental Physics, Slovak Academy of Sciences, Ko\v{s}ice, Slovak Republic\\
$^{59}$ Institute of Physics, Homi Bhabha National Institute, Bhubaneswar, India\\
$^{60}$ Institute of Physics of the Czech Academy of Sciences, Prague, Czech Republic\\
$^{61}$ Institute of Space Science (ISS), Bucharest, Romania\\
$^{62}$ Institut f\"{u}r Kernphysik, Johann Wolfgang Goethe-Universit\"{a}t Frankfurt, Frankfurt, Germany\\
$^{63}$ Instituto de Ciencias Nucleares, Universidad Nacional Aut\'{o}noma de M\'{e}xico, Mexico City, Mexico\\
$^{64}$ Instituto de F\'{i}sica, Universidade Federal do Rio Grande do Sul (UFRGS), Porto Alegre, Brazil\\
$^{65}$ Instituto de F\'{\i}sica, Universidad Nacional Aut\'{o}noma de M\'{e}xico, Mexico City, Mexico\\
$^{66}$ iThemba LABS, National Research Foundation, Somerset West, South Africa\\
$^{67}$ Jeonbuk National University, Jeonju, Republic of Korea\\
$^{68}$ Johann-Wolfgang-Goethe Universit\"{a}t Frankfurt Institut f\"{u}r Informatik, Fachbereich Informatik und Mathematik, Frankfurt, Germany\\
$^{69}$ Korea Institute of Science and Technology Information, Daejeon, Republic of Korea\\
$^{70}$ KTO Karatay University, Konya, Turkey\\
$^{71}$ Laboratoire de Physique des 2 Infinis, Ir\`{e}ne Joliot-Curie, Orsay, France\\
$^{72}$ Laboratoire de Physique Subatomique et de Cosmologie, Universit\'{e} Grenoble-Alpes, CNRS-IN2P3, Grenoble, France\\
$^{73}$ Lawrence Berkeley National Laboratory, Berkeley, California, United States\\
$^{74}$ Lund University Department of Physics, Division of Particle Physics, Lund, Sweden\\
$^{75}$ Nagasaki Institute of Applied Science, Nagasaki, Japan\\
$^{76}$ Nara Women{'}s University (NWU), Nara, Japan\\
$^{77}$ National and Kapodistrian University of Athens, School of Science, Department of Physics , Athens, Greece\\
$^{78}$ National Centre for Nuclear Research, Warsaw, Poland\\
$^{79}$ National Institute of Science Education and Research, Homi Bhabha National Institute, Jatni, India\\
$^{80}$ National Nuclear Research Center, Baku, Azerbaijan\\
$^{81}$ National Research and Innovation Agency - BRIN, Jakarta, Indonesia\\
$^{82}$ Niels Bohr Institute, University of Copenhagen, Copenhagen, Denmark\\
$^{83}$ Nikhef, National institute for subatomic physics, Amsterdam, Netherlands\\
$^{84}$ Nuclear Physics Group, STFC Daresbury Laboratory, Daresbury, United Kingdom\\
$^{85}$ Nuclear Physics Institute of the Czech Academy of Sciences, Husinec-\v{R}e\v{z}, Czech Republic\\
$^{86}$ Oak Ridge National Laboratory, Oak Ridge, Tennessee, United States\\
$^{87}$ Ohio State University, Columbus, Ohio, United States\\
$^{88}$ Physics department, Faculty of science, University of Zagreb, Zagreb, Croatia\\
$^{89}$ Physics Department, Panjab University, Chandigarh, India\\
$^{90}$ Physics Department, University of Jammu, Jammu, India\\
$^{91}$ Physics Department, University of Rajasthan, Jaipur, India\\
$^{92}$ Physics Program and International Institute for Sustainability with Knotted Chiral Meta Matter (SKCM2), Hiroshima University, Hiroshima, Japan\\
$^{93}$ Physikalisches Institut, Eberhard-Karls-Universit\"{a}t T\"{u}bingen, T\"{u}bingen, Germany\\
$^{94}$ Physikalisches Institut, Ruprecht-Karls-Universit\"{a}t Heidelberg, Heidelberg, Germany\\
$^{95}$ Physik Department, Technische Universit\"{a}t M\"{u}nchen, Munich, Germany\\
$^{96}$ Politecnico di Bari and Sezione INFN, Bari, Italy\\
$^{97}$ Research Division and ExtreMe Matter Institute EMMI, GSI Helmholtzzentrum f\"ur Schwerionenforschung GmbH, Darmstadt, Germany\\
$^{98}$ Saha Institute of Nuclear Physics, Homi Bhabha National Institute, Kolkata, India\\
$^{99}$ School of Physics and Astronomy, University of Birmingham, Birmingham, United Kingdom\\
$^{100}$ Secci\'{o}n F\'{\i}sica, Departamento de Ciencias, Pontificia Universidad Cat\'{o}lica del Per\'{u}, Lima, Peru\\
$^{101}$ Stefan Meyer Institut f\"{u}r Subatomare Physik (SMI), Vienna, Austria\\
$^{102}$ SUBATECH, IMT Atlantique, Nantes Universit\'{e}, CNRS-IN2P3, Nantes, France\\
$^{103}$ Suranaree University of Technology, Nakhon Ratchasima, Thailand\\
$^{104}$ Technical University of Ko\v{s}ice, Ko\v{s}ice, Slovak Republic\\
$^{105}$ The Henryk Niewodniczanski Institute of Nuclear Physics, Polish Academy of Sciences, Cracow, Poland\\
$^{106}$ The University of Texas at Austin, Austin, Texas, United States\\
$^{107}$ Universidad Aut\'{o}noma de Sinaloa, Culiac\'{a}n, Mexico\\
$^{108}$ Universidade de S\~{a}o Paulo (USP), S\~{a}o Paulo, Brazil\\
$^{109}$ Universidade Estadual de Campinas (UNICAMP), Campinas, Brazil\\
$^{110}$ Universidade Federal do ABC, Santo Andre, Brazil\\
$^{111}$ University of Cape Town, Cape Town, South Africa\\
$^{112}$ University of Houston, Houston, Texas, United States\\
$^{113}$ University of Jyv\"{a}skyl\"{a}, Jyv\"{a}skyl\"{a}, Finland\\
$^{114}$ University of Kansas, Lawrence, Kansas, United States\\
$^{115}$ University of Liverpool, Liverpool, United Kingdom\\
$^{116}$ University of Science and Technology of China, Hefei, China\\
$^{117}$ University of South-Eastern Norway, Kongsberg, Norway\\
$^{118}$ University of Tennessee, Knoxville, Tennessee, United States\\
$^{119}$ University of the Witwatersrand, Johannesburg, South Africa\\
$^{120}$ University of Tokyo, Tokyo, Japan\\
$^{121}$ University of Tsukuba, Tsukuba, Japan\\
$^{122}$ University Politehnica of Bucharest, Bucharest, Romania\\
$^{123}$ Universit\'{e} Clermont Auvergne, CNRS/IN2P3, LPC, Clermont-Ferrand, France\\
$^{124}$ Universit\'{e} de Lyon, CNRS/IN2P3, Institut de Physique des 2 Infinis de Lyon, Lyon, France\\
$^{125}$ Universit\'{e} de Strasbourg, CNRS, IPHC UMR 7178, F-67000 Strasbourg, France, Strasbourg, France\\
$^{126}$ Universit\'{e} Paris-Saclay Centre d'Etudes de Saclay (CEA), IRFU, D\'{e}partment de Physique Nucl\'{e}aire (DPhN), Saclay, France\\
$^{127}$ Universit\`{a} degli Studi di Foggia, Foggia, Italy\\
$^{128}$ Universit\`{a} del Piemonte Orientale, Vercelli, Italy\\
$^{129}$ Universit\`{a} di Brescia, Brescia, Italy\\
$^{130}$ Variable Energy Cyclotron Centre, Homi Bhabha National Institute, Kolkata, India\\
$^{131}$ Warsaw University of Technology, Warsaw, Poland\\
$^{132}$ Wayne State University, Detroit, Michigan, United States\\
$^{133}$ Westf\"{a}lische Wilhelms-Universit\"{a}t M\"{u}nster, Institut f\"{u}r Kernphysik, M\"{u}nster, Germany\\
$^{134}$ Wigner Research Centre for Physics, Budapest, Hungary\\
$^{135}$ Yale University, New Haven, Connecticut, United States\\
$^{136}$ Yonsei University, Seoul, Republic of Korea\\
$^{137}$  Zentrum  f\"{u}r Technologie und Transfer (ZTT), Worms, Germany\\
$^{138}$ Affiliated with an institute covered by a cooperation agreement with CERN\\
$^{139}$ Affiliated with an international laboratory covered by a cooperation agreement with CERN.\\

\end{flushleft}

\end{document}